\documentclass[11pt,english,conference]{IEEEtran}
\usepackage[T1]{fontenc}
\usepackage[latin9]{inputenc}
\usepackage{verbatim}
\usepackage{amsmath}
\usepackage{graphicx}
\usepackage{amssymb}
\usepackage{esint}



\usepackage{amsthm}\usepackage{dsfont}\usepackage{array}\usepackage{mathrsfs}\usepackage{cite}\usepackage{comment}\usepackage{mathrsfs}\usepackage{hyperref}

\makeatother

\usepackage{babel}

\makeatother

\usepackage{babel}

\makeatother

\usepackage{babel}

\begin{document}
\bibliographystyle{IEEEtran}

\title{Channel Capacity under \\ General Nonuniform Sampling}

\author{\IEEEauthorblockN{Yuxin Chen} \IEEEauthorblockA{EE, Stanford
University\\
 Email: yxchen@stanford.edu} \and \IEEEauthorblockN{Yonina
C. Eldar} \IEEEauthorblockA{EE, Technion\\
 Email: yonina@ee.technion.ac.il} \and \IEEEauthorblockN{Andrea
J. Goldsmith} \IEEEauthorblockA{EE, Stanford University\\
 Email: andrea@ee.stanford.edu}}
\maketitle
\begin{abstract}
This paper develops the fundamental capacity limits of a sampled analog
channel under a sub-Nyquist sampling rate constraint. In particular,
we derive the capacity of sampled analog channels over a general class
of time-preserving sampling methods including irregular nonuniform
sampling. Our results indicate that the optimal sampling structures
extract out the set of frequencies that exhibits the highest SNR among
all spectral sets of support size equal to the sampling rate. The
capacity under sub-Nyquist sampling can be attained through filter-bank
sampling, or through a single branch of modulation and filtering followed
by uniform sampling. The capacity under sub-Nyquist sampling is a
monotone function of the sampling rate. These results indicate that
the optimal sampling schemes suppress aliasing, and that employing
irregular nonuniform sampling does not provide capacity gain over
uniform sampling sets with appropriate preprocessing for a large class
of channels.\end{abstract}
\theoremstyle{plain}\newtheorem{lem}{\textbf{Lemma}}\newtheorem{theorem}{\textbf{Theorem}}\newtheorem{corollary}{\textbf{Corollary}}\newtheorem{prop}{\textbf{Proposition}}\newtheorem{fct}{Fact}\newtheorem{remark}{\textbf{Remark}}

\theoremstyle{definition}\newtheorem{definition}{\textbf{Definition}}\newtheorem{example}{\textbf{Example}}

\section{Introduction}

Capacity of analog channels along with the capacity-achieving transmission
strategies was pioneered by Shannon. These results have provided fundamental
insights for modern communication system design. Most Shannon capacity
results (e.g. \cite{Gallager68,Med2000}) focus on the analog capacity
commensurate with sampling at or above twice the channel bandwidth,
which does not explicitly account for the effects upon capacity of
sub-Nyquist rate sampling. In practice, however, hardware and power
limitations may preclude sampling at the Nyquist rate associated with
the channel bandwidth. On the other hand, although the Nyquist sampling
rate is necessary for perfect recovery of bandlimited functions, this
rate can be excessive when certain signal structures are properly
exploited. Inspired by recent {}``compressive sensing'' ideas, sub-Nyquist
sampling approaches have been developed to exploit the structure of
various classes of input signals with different structures (e.g. \cite{MisEld2010Theory2Practice}).

Although optimal sampling methods have been extensively explored in
the sampling literature, they are typically investigated either under
a noiseless setting, or based on statistical reconstruction measures
(e.g. mean squared error (MSE)). Berger\emph{ et. al. }\cite{BergerThesis}
related MSE-based optimal sampling with capacity for several special
channels but did not derive the sampled capacity for more general
channels. Our recent work \cite{ChenGolEld2010} established a new
framework that characterized sub-Nyquist sampled channel capacity
for a broad class of sampling methods, including filter-bank and modulation-bank
sampling \cite{Papoulis1977,MisEld2010Theory2Practice}. For these
sampling methods, we determined optimal sampling structures based
on capacity as a metric, illuminated intriguing connections between
MIMO channel capacity and capacity of undersampled channels, as well
as a new connection between capacity and MSE.

One interesting fact we discovered in this previous work is the non-monotonicity
of capacity with sampling rate under filter- and modulation-bank sampling,
assuming an equal sampling rate per branch for a given number of branches.
This indicates that more sophisticated sampling schemes, adaptive
to the sampling rate, are needed to maximize capacity under sub-Nyquist
rate constraints, including both uniform and nonuniform sampling.
Beurling pioneered the investigation of general nonuniform sampling
for bandlimited functions. However, it is unclear which sampling method
can best exploit the channel structure, thereby maximizing sampled
capacity under a sub-Nyquist sampling rate constraint. Although several
classes of sampling methods were shown in \cite{ChenGolEld2010} to
have a closed-form capacity solution, the capacity limit might not
exist for general sampling. It remains unknown whether there exists
a capacity upper bound over a general class of sub-Nyquist sampling
systems and, if so, when the bound is achievable. 

In this paper, we derive the sub-Nyquist sampled channel capacity
for a general class of time-preserving nonuniform sampling methods.
We demonstrate that the fundamental limit can be achieved through
filter-bank sampling with varied sampling rate at different branches,
or a single branch of modulation and filtering followed by uniform
sampling. Our results indicate that irregular sampling sets, which
are more complicated to realize in hardware, do not provide capacity
increase compared with regular uniform sampling sets for a broad class
of channels. Furthermore, we demonstrate that the optimal sampling
schemes suppress aliasing through filter bank, modulation, or input
optimization.

{}

\section{Sampled Channel Capacity}

\subsection{System Model}

We consider a waveform channel, which is modeled as a linear time-invariant
(LTI) filter with impulse response $h(t)$ and frequency response
$H(f)=\int_{-\infty}^{\infty}h(t)e^{-j2\pi ft}\text{d}t$. The channel
output is given by\begin{equation}
r(t)=h(t)*x(t)+\eta(t),\label{eq:ChannelModel}\end{equation}
 where $x(t)$ is the transmitted signal, and $\eta(t)$ is stationary
Gaussian noise with power spectral density $\mathcal{S}_{\eta}\left(f\right)$.
We assume throughout that \textit{perfect channel state information}
is known at both the transmitter and receiver.

\begin{figure}[htbp]
\begin{centering}
\textsf{\includegraphics[scale=0.3]{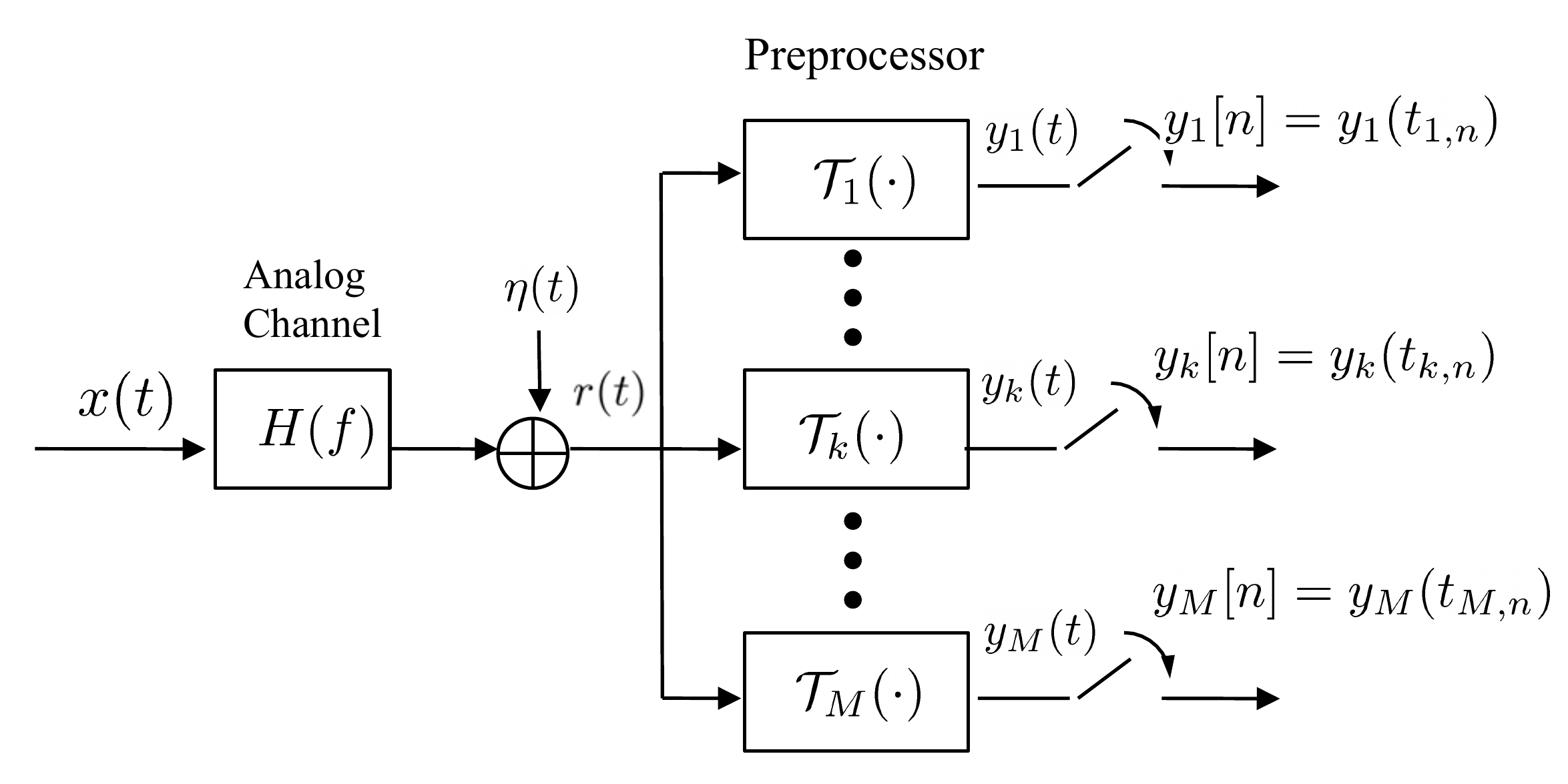}} 
\par\end{centering}

\caption{\label{fig:ProblemFormulation} The input $x(t)$ is constrained to
$[-T,T]$ and passed through an analog channel and contaminated by
noise $\eta(t)$. The analog channel output $r(t)$ is then passed
through a linear preprocessing system $\mathcal{T}$. The preprocessed
output $y(t)$ is observed over $[-T,T]$ and sampled on the sampling
set $\Lambda=\left\{ t_{n}\mid n\in\mathbb{Z}\right\} $.}
\end{figure}

The analog channel output is passed through $M$ ($1\leq M\leq\infty$)
linear preprocessing systems each followed by a pointwise sampler,
as illustrated in Fig. \ref{fig:ProblemFormulation}. The preprocessed
output $y_{k}(t)$ at the $k$th branch is obtained by applying a
linear operator $\mathcal{T}_{k}$ to $r(t)$, i.e. $y_{k}(t)=\mathcal{T}_{k}\left(r(t)\right)$.
The linear operators can be time-varying, and include filtering and
modulation as special cases. We define the impulse response $q(t,\tau)$
of a time-varying system as the output seen at time $t$ due to an
impulse in the input at time $\tau$. The pointwise sampler following
the preprocessor can be \emph{nonuniform}. The preprocessed output
$y_{k}(t)$ is sampled at times $t_{k,n}\left(n\in\mathbb{Z}\right)$,
yielding a sequence $ $$y_{k}[n]=y_{k}\left(t_{k,n}\right).$ At
the $k$th branch, the\emph{ sampling set} is defined by $\Lambda_{k}:=\left\{ t_{k,n}\mid n\in\mathbb{Z}\right\} .$
When $t_{k,n}=nT_{s}$, $\Lambda_{k}$ is said to be uniform with
period $T_{s}$.

\subsection{Sampling Rate}

In general, the sampling set $\Lambda$ may be irregular. This calls
for a generalized definition of the sampling rate. One notion commonly
used in sampling theory is the Beurling density \cite{AldGro2001}
as defined below.

\begin{definition}[\bf Beurling Density]For a sampling set $\Lambda$,
the upper and lower Beurling density are defined as\[
\begin{cases}
D^{+}\left(\Lambda\right) & =\lim_{r\rightarrow\infty}\sup_{z\in\mathbb{R}}\frac{\text{cardinality}\left(\Lambda\cap\left[z,z+r\right]\right)}{r},\\
D^{-}\left(\Lambda\right) & =\lim_{r\rightarrow\infty}\inf_{z\in\mathbb{R}}\frac{\text{cardinality}\left(\Lambda\cap\left[z,z+r\right]\right)}{r}.\end{cases}\]
 When $D^{+}\left(\Lambda\right)=D^{-}\left(\Lambda\right)$, the
sampling set $\Lambda$ is said to be of uniform Beurling density
$D\left(\Lambda\right):=D^{-}\left(\Lambda\right)$.\end{definition}When
the sampling set is uniform with period $T_{s}$, the Beurling density
is $D(\Lambda)=1/T_{s}$, which coincides with our conventional definition
of the sampling rate.

%
{}

Given a preprocessed output $y_{k}(t)$, we can use Beurling density
to characterize the sampling rate on $y_{k}(t)$. However, since the
preprocessor might distort the time scale of the input, the resulting
{}``sampling rate'' might not make physical sense, as illustrated
below.

\begin{example}[Compressor]Consider a preprocessor defined by the
relation $y(t)=\mathcal{T}\left(r(t)\right)=r\left(Lt\right)$ with
$L\geq2$ being a positive integer. If we apply a uniform sampling
set $ $$\Lambda=\{t_{n}:t_{n}=n/f_{s}\}$ on $y(t)$, then the sampled
sequence at a {}``sampling rate'' $f_{s}$ is given by $y[n]=y\left(n/f_{s}\right)=r\left(nL/f_{s}\right)$,
which corresponds to sampling $r(t)$ at rate $f_{s}/L$. The compressor
effectively time-warps the signal, thus resulting in a mismatch of
the time scales between the input and output.\end{example}

The example of the compressor illustrates that the notion of a given
sampling rate may be misleading for systems that exhibit time warping.
Hence, our results will focus on sampling that preserves time scales.
A class of linear systems that preserves time scales are modulation
operators $\left(y(t)=p(t)x(t),\forall t\right)$, which perform pointwise
scaling of the input, and hence do not change the time scale. Another
class are periodic systems which includes LTI filtering, and are defined
as follows. 

\begin{definition}[\bf Periodic System]A linear preprocessing system
is said to be periodic with period $T_{q}$ if its impulse response
$q(t,\tau)$ satisfies \begin{equation}
q(t,\tau)=q(t+T_{q},\tau+T_{q}),\quad\forall t,\tau\in\mathbb{R}.\label{eq:ShiftInvariantImpulseResponse}\end{equation}
 \end{definition}

A more general class of systems that preserve the time scale can be
generated through modulation and periodic subsystems. Specifically,
we can define a general time-preserving system by connecting a set
of modulation or periodic operators in parallel or in serial. This
leads to the following definition.

\begin{definition}[\bf Time-preserving System]Given an index set
$\mathcal{I}$, a preprocessing system $\mathcal{T}:x(t)\mapsto\left\{ y_{k}(t),k\in\mathcal{I}\right\} $
is said to be time-preserving if

(1) The system input is passed through $\left|\mathcal{I}\right|$
(possibly countably many) branches of linear preprocessors, yielding
a set of analog outputs $\left\{ y_{k}(t)\mid k\in\mathcal{I}\right\} $.

(2) In each branch, the preprocessor comprises a set of periodic or
modulation operators connected in serial.

\end{definition}

With a preprocessing system that preserves the time scale, we can
now define the aggregate sampling rate through Beurling density.

\begin{definition}A sampling system is said to be time-preserving
with sampling rate $f_{s}$ if

(1) Its preprocessing system $\mathcal{T}$ is time-preserving.

(2) The preprocessed output $y_{k}(t)$ is sampled by a sampling set
$\Lambda_{k}=\left\{ t_{l,k}\mid l\in\mathbb{Z}\right\} $ with a
uniform Beurling density $f_{k,\text{s}}$, which satisfies $\sum_{k\in\mathcal{I}}f_{k,\text{s}}=f_{s}$.

\end{definition}

%
{}

\subsection{Capacity Definition}

Suppose that the transmit signal $x(t)$ is constrained to the time
interval $\left[-T,T\right]$, and the received signal $y(t)$ is
sampled and observed over $\left[-T,T\right]$. For a given sampling
system $\mathcal{P}$ that consists of a preprocessor $\mathcal{T}$
and a sampling set $\Lambda$, and a given time duration $T$, the
capacity $C_{T}^{\mathcal{P}}(f_{s},P)$ is defined as \[
C_{T}^{\mathcal{P}}(f_{s},P)=\max_{p(x)}\frac{1}{2T}I\left(x\left(\left[-T,T\right]\right),\left\{ y[n]\right\} _{\left[-T,T\right]}\right)\]
 subject to a power constraint $\mathbb{E}(\frac{1}{2T}\int_{-T}^{T}|x(t)|^{2}\mathrm{d}t)\leq P$.
Here, $\left\{ y[t_{n}]\right\} _{\left[-T,T\right]}$ denotes the
set of samples obtained within time $\left[-T,T\right]$. The sub-Nyquist
sampled channel capacity for the given system can be studied by taking
the limit as $T\rightarrow\infty$. It was shown in \cite{ChenGolEld2010}
that $\lim_{T\rightarrow\infty}C_{T}^{\mathcal{P}}(f_{s},P)$ exists
for a broad class of sampling methods. We caution, however, that the
existence of the limit is not guaranteed for all sampling methods,
e.g. the limit might not exist for an irregular sampling set. The
capacity and an upper bound under general nonuniform sampling is defined
as follows.

\begin{definition}(1) $C^{\mathcal{P}}(P)$ is said to be the \emph{capacity}
of a given sampled analog channel if $\lim_{T\rightarrow\infty}C_{T}^{\mathcal{P}}(f_{s},P)$
exists and $C^{\mathcal{P}}(f_{s},P)=\lim_{T\rightarrow\infty}C_{T}^{\mathcal{P}}(f_{s},P)$;

(2) $C_{\text{u}}^{\mathcal{P}}(P)$ is said to be a \emph{capacity
upper bound} of the sampled channel if $C_{\text{u}}^{\mathcal{P}}(f_{s},P)\geq\lim\sup_{T\rightarrow\infty}C_{T}^{\mathcal{P}}(f_{s},P)$.\end{definition}

The above capacity is defined for a specific sampling system. Another
metric of interest is the maximum date rate for all sampling schemes
within a general class of nonuniform sampling systems. This motivates
us to define the sub-Nyquist sampled channel capacity for the class
of linear time-preserving systems as follows.

\begin{definition}[\bf Sampled Capacity under Time-preserving Linear Sampling](1)
$C(f_{s},P)$ is said to be the \emph{sampled capacity} of an analog
channel under time-preserving linear sampling for a given sampling
rate $f_{s}$ if $C(f_{s},P)=\sup_{\mathcal{P}}C^{\mathcal{P}}(f_{s},P)$;

(2) $C_{\text{u}}(P)$ is said to be a \emph{capacity upper bound}
of the sampled channel under this sampling if $C_{\text{u}}(f_{s},P)\geq\sup_{\mathcal{P}}\lim\sup_{T\rightarrow\infty}C_{T}^{\mathcal{P}}(f_{s},P)$.

Here, the supremum on $\mathcal{P}$ is over all time-preserving linear
sampling systems.\end{definition}

\section{Capacity Analysis}

\subsection{An Upper Bound on Sampled Channel Capacity}

A time-preserving sampling system preserves the time scale of the
signal, and hence does not compress or expand the frequency response.
We now determine an upper limit on the sampled channel capacity for
this class of general nonuniform sampling systems.

\begin{theorem}[\bf Converse]\label{thm:GeneralSampledCapacity}Consider
a time-preserving sampling system with sampling rate $f_{s}$. Suppose
that the output impulse response of the sampled channel is of finite
duration, and that there exists a frequency set $B_{\mathrm{m}}$
that satisfies $\mu\left(B_{\mathrm{m}}\right)=f_{s}$ and\[
{\displaystyle \int}_{f\in B_{\mathrm{m}}}\frac{\left|H(f)\right|^{2}}{\mathcal{S}_{\eta}(f)}\mathrm{d}f=\sup_{B:\mu\left(B\right)=f_{s}}{\displaystyle \int}_{f\in B}\frac{\left|H(f)\right|^{2}}{\mathcal{S}_{\eta}(f)}\mathrm{d}f,\]
 where $\mu\left(\cdot\right)$ denotes the Lebesgue measure. Then
the sampled channel capacity is upper bounded by\begin{equation}
C_{\mathrm{u}}\left(f_{s},P\right)={\displaystyle \int}_{f\in B_{\mathrm{m}}}\frac{1}{2}\left[\log\left(\nu\frac{\left|H(f)\right|^{2}}{\mathcal{S}_{\eta}(f)}\right)\right]^{+}\mathrm{d}f,\label{eq:GeneralCapacityUpperBound}\end{equation}
 where $[x]^{+}\overset{\Delta}{=}\max\left(x,0\right)$ and $\nu$
satisfies\begin{align}
{\displaystyle \int}_{f\in B_{\mathrm{m}}}\left[\nu-\frac{\left|H(f)\right|^{2}}{\mathcal{S}_{\eta}(f)}\right]^{+}\mathrm{d}f & =P.\label{eq:GeneralCapacityWaterLevel}\end{align}

\end{theorem}

In other words, the upper limit is equivalent to the maximum capacity
of a channel whose spectral occupancy is no larger than $f_{s}$.
The above result basically implies that even if we allow for more
complex irregular sampling sets, the sampled capacity cannot exceed
the one commensurate with the analog capacity when constraining all
transmit signals to the interval of bandwidth $f_{s}$ that experience
the highest SNR. Accordingly, the optimal input distribution will
lie in this frequency set. This theorem also indicates that the capacity
is attained when aliasing is suppressed by the sampling structure,
as will be seen later in our capacity-achieving scheme. When the optimal
frequency interval $B_{\text{m}}$ is selected, a water filling power
allocation strategy is performed over the spectral domain with water
level $\nu$.

This theorem can be approximately interpreted based on a Fourier domain
analysis. The Fourier transform of the analog channel output is given
by $H(f)X(f)+N(f)$, where $X(f)$ and $N(f)$ denote, respectively,
the Fourier response of $x(t)$ and $\eta(t)$. This output is passed
through the sampling system to yield a sequence at a rate $f_{s}$,
which can be further mapped to the space of bandlimited functions
$\mathcal{L}_{2}(-f_{s}/2,f_{s}/2)$ through linear mapping without
frequency warping. The whitening operation for the noise component,
combined with the sampling system operator, forms an orthonormal mapping
from $\mathcal{L}_{2}(-\infty,\infty)$ to $\mathcal{L}_{2}(-f_{s}/2,f_{s}/2)$.
The optimal orthonormal mapping that maximizes SNR is to extract out
a frequency set $B_{\text{m}}$ of size $f_{s}$ that contains the
frequency components with the highest SNR, which leads to the capacity
upper bound (\ref{eq:GeneralCapacityUpperBound}).

{}

The outline of the proof of Theorem \ref{thm:GeneralSampledCapacity}
is sketched below. We start from the capacity of periodic sampling
whose sampled channel capacity exists, and then derive the upper bound
through finite-duration approximation of the true channels. Details
can be found in \cite{ChenEldarGoldsmith2012}.

\begin{figure}[htbp]
\begin{centering}
\textsf{\includegraphics[scale=0.27]{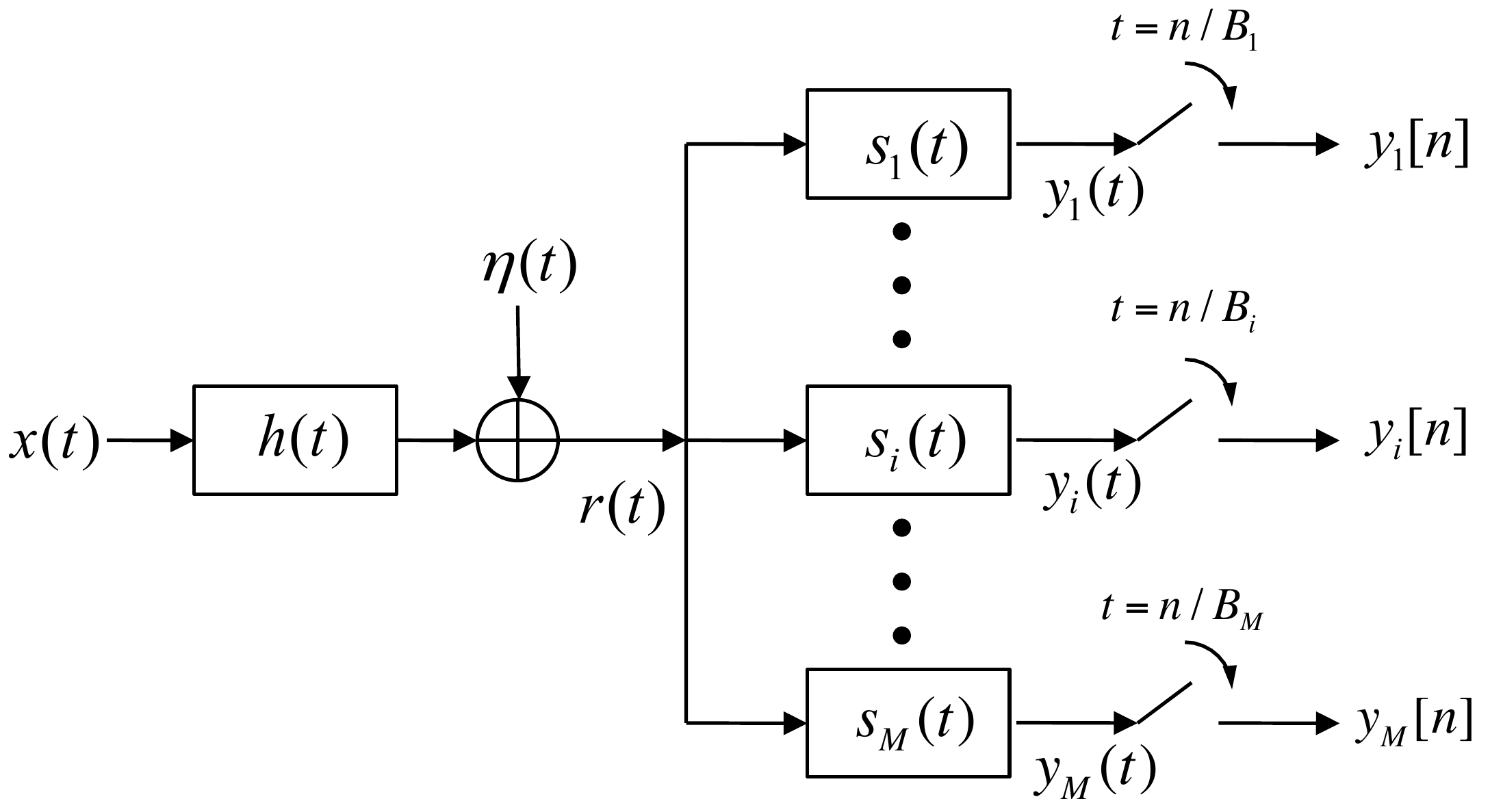}} 
\par\end{centering}

\begin{centering}
(a) 
\par\end{centering}

\begin{centering}
\textsf{\includegraphics[scale=0.27]{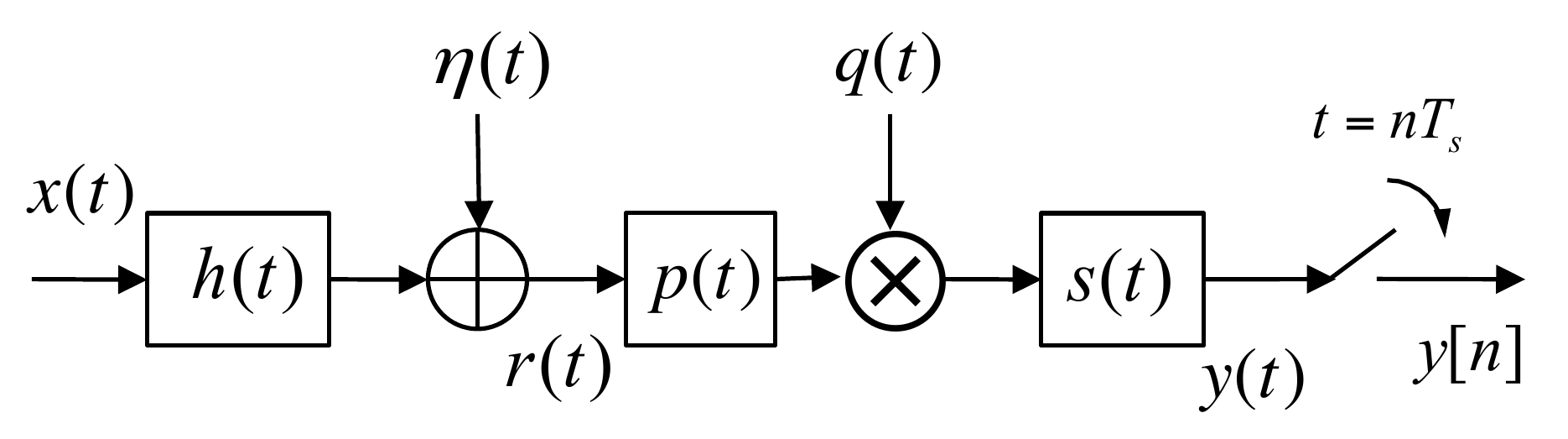}} 
\par\end{centering}

\begin{centering}
(b) 
\par\end{centering}

\caption{\label{fig:FilterBankModulation} (a) Filter-bank sampling: each branch
filters out a frequency interval of bandwidth $B_{k}$, and samples
it with rate $f_{k,\text{s}}=B_{k}$; (b) A single branch of modulation
and filtering: the channel output is prefiltered by a filter with
impulse response $p(t)$, modulated by a sequence $q(t)$, post-filtered
by another filter of impulse response $s(t)$, and finally sampled
uniformly at a rate $f_{s}$.}
\end{figure}

Suppose first that the whole sampling system is periodic, where the
impulse response $q(t,\tau)$ is periodic with period $T_{q}$ ($f_{s}T_{q}\in\mathbb{Z}$)
and the sampling set obeys $t_{k+f_{s}T_{q}}=t_{k}+T_{q},\forall k\in\mathbb{Z}$.
The periodicity of the system guarantees the existence of $\lim_{T\rightarrow\infty}C_{T}^{\mathcal{P}}$.
Specifically, denote by $Q_{k}(f)$ the Fourier transform $Q_{k}(f):=\int_{-\infty}^{\infty}q(t_{k},t_{k}-t)\exp(-j2\pi ft)\mathrm{d}t$,
and introduce an $f_{q}T_{s}\times\infty$ dimensional matrix ${\bf F}_{q}\left(f\right)$
and an infinite diagonal square matrix ${\bf F}_{h}\left(f\right)$
such that for all $m,l\in\mathbb{Z}$ and $1\leq k\leq f_{q}T_{s}$,\[
\left({\bf F}_{q}\right)_{k,l}\left(f\right):=Q_{k}\left(f+lf_{q}\right),\quad\left({\bf F}_{h}\right)_{l,l}(f)=H\left(f+lf_{q}\right).\]
We can then express in closed form the sampled analog capacity as
given in the following theorem.

\begin{theorem}[\bf Capacity for Periodic Sampling]\label{thm:ShiftInvariantSampledCapacity}Suppose
the sampling system $\mathcal{P}$ is periodic with period $T_{q}=1/f_{q}$
and sampling rate $f_{s}$. Assume that $\left|H(f)Q_{k}(f)\right|^{2}/\mathcal{S}_{\eta}(f)$
is bounded and satisfies $\int_{-\infty}^{\infty}\left|H(f)Q_{k}(f)\right|^{2}/\mathcal{S}_{\eta}(f)<\infty$
for all $1\leq k\leq f_{q}T_{s}$, and define ${\bf F}_{w}=\left({\bf F}_{q}{\bf F}_{q}^{*}\right)^{-\frac{1}{2}}{\bf F}_{q}{\bf F}_{h}$.
Then\[
C^{\mathcal{P}}\left(f_{s},P\right)=\frac{1}{2}{\displaystyle \int}_{-f_{q}/2}^{f_{q}/2}\sum_{i=1}^{f_{s}T_{q}}\left[\log\left(\nu\lambda_{i}\left\{ {\bf F}_{w}{\bf F}_{w}^{*}\right\} \right)\right]^{+}\mathrm{d}f,\]
 where $\nu$ is chosen according to the water-filling strategy.

\end{theorem}

We observe that the capacity of any periodic sampling system cannot
exceed the capacity (\ref{eq:GeneralCapacityUpperBound}). 

Now we consider the more general sampling system that might not be
periodic. For a given input and output duration $[-T,T]$, the impulse
response $h(t,\tau)$ $\left(|t|,|\tau|\leq T\right)$ can be extended
periodically to generate an impulse response of a periodic system.
Suppose first that the impulse response is of finite duration, then
for sufficiently large $T$, the sampled capacity $C_{T}$ can be
upper bounded arbitrarily closely by the capacity of the generated
periodic system, which are further bounded by the upper limit (\ref{eq:GeneralCapacityUpperBound}).
Since the impulse response is constrained in $\mathcal{L}^{2}$ space,
the leakage signal between different blocks can be made arbitrarily
weak by introducing a guard zone with length $T^{1-\epsilon}$. This
shows the full generality of our upper bound.

\subsection{Achievability}

For most scenarios of physical interest, the capacity upper bound
given in Theorem \ref{thm:GeneralSampledCapacity} can be achieved
through filter-bank sampling. 

\begin{theorem}[\bf Achievability]\label{cor:OptimalSamplingGeneralSampledCapacity}Suppose
that the SNR $|H(f)|^{2}/\mathcal{S}_{\eta}(f)$ of the analog channel
is continuous and Riemann integrable. Then the maximizing frequency
set $B_{\mathrm{m}}$ defined in Theorem \ref{thm:GeneralSampledCapacity}
can be divided into $B_{\mathrm{m}}=\cup_{i}B_{i}\text{ }\cup\text{ }D,$
where $D$ contains a set of singular points, $B_{i}$ is a continuous
interval, and $D$ and $B_{i}(i\in\mathbb{N})$ are non-overlapping
sets. The upper bound in (\ref{eq:GeneralCapacityUpperBound}) can
be achieved by filter-bank sampling. Specifically, in the $k^{\text{th}}$
branch, the frequency response of the filter is given by\[
S_{k}(f)=\begin{cases}
1,\quad & \text{if }f\in B_{k},\\
0, & \text{otherwise},\end{cases}\]
 and the filter is followed by a uniform sampler with sampling rate
$\mu\left(B_{k}\right)$.

\end{theorem}

Since the bandwidth of $B_{i}$ may be irrational and the system may
require an infinite number of filters, the sampling system is in general
aperiodic. However, filter-bank sampling with \emph{varied sampling
rates} in different branches outperforms all other sampling mechanisms
in maximizing capacity. $ $

The optimality of filter-bank sampling immediately leads to another
optimal sampling structure. As we have shown in \cite{ChenGolEld2010},
filter-bank sampling can be replaced by a single branch of modulation
and filtering as illustrated in Fig. \ref{fig:FilterBankModulation},
which can approach the capacity arbitrarily closely if the spectral
support can be divided into subbands with constant SNR. A channel
of physical interest can often be approximated as piecewise constant
in this way. Given the maximizing frequency set $B_{\text{m}}$, we
first suppress the frequency components outside $B_{\text{m}}$ using
an LTI prefilter. A modulation module is then applied to move all
frequency components within $B_{\text{m}}$ to the baseband $[-f_{s}/2,f_{s}/2]$.
The aliasing effect can be significantly mitigated by appropriate
choices of modulation weights for different spectral subbands. We
then employ another low-pass filter to suppress out-of-band signals,
and sample the output using a pointwise uniform sampler. The optimizing
modulation sequence can be found in \cite{ChenGolEld2010,ChenEldarGoldsmith2012}.
Compared with filter-bank sampling, a single branch of modulation
and filtering only requires the design of a low-pass filter, a band-pass
filter and a multiplication module, which are typically lower complexity
to implement than a filter bank.

\section{Discussion}

The above analytical results characterize the sampled capacity for
a general class of sampling methods. Some properties of the capacity
results are as follows.

\textbf{Monotonicity}. It can be seen from (\ref{eq:GeneralCapacityUpperBound})
that increasing the sampling rate from $f_{s}$ to $\tilde{f_{s}}$
requires us to crop out another frequency set $\tilde{B}_{\text{m}}$
of size $\tilde{f}_{s}$ that has the highest SNRs. The original frequency
set $B_{\text{m}}$ we choose must be a subset of $\tilde{B}_{\text{m}}$,
and hence the sampled capacity with rate $\tilde{f}_{s}$ is no lower
than that with rate $f_{s}$.

\textbf{Irregular sampling} \textbf{set}. Sampling with irregular
sampling sets, while requiring complicated reconstruction techniques
\cite{AldGro2001}, does not outperform filter-bank or modulation-bank
sampling with regular uniform sampling sets in maximizing achievable
data rate. 

\textbf{Alias suppression}. Aliasing does not allow a higher capacity
to be achieved. The optimal sampling method corresponds to the optimal
alias-suppression strategy. This is in contrast to the benefits obtained
through scrambling of spectral contents in many sub-Nyquist sampling
schemes with unknown signal supports.

\textbf{Perturbation of sampling} \textbf{set}. If the optimal filter-bank
or modulation sampling is employed, mild perturbation of post-filtering
uniform sampling sets does not degrade the sampled capacity. For example,
suppose that a sampling rate $\hat{f}_{s}$ is used in any branch
and the sampling set satisfies $\left|\hat{t}_{n}-n/\hat{f}_{s}\right|\leq\hat{f}_{s}/4$.
Kadec has shown that $\left\{ \exp\left(j2\pi\hat{t}_{n}f\right)\mid n\in\mathbb{Z}\right\} $
also forms a Riesz basis of $\mathcal{L}^{2}(-\hat{f}_{s}/2,\hat{f}_{s}/2)$,
thereby preserving information integrity. The sampled capacity is
invariant under mild perturbation of the sampling sets.

\textbf{Hardware implementation}. When the sampling rate is increased
from $f_{s1}$ to $f_{s2}$, we need only to insert an additional
filter bank of overall sampling rate $f_{s2}-f_{s1}$ to select another
set of spectral components with bandwidth $f_{s2}-f_{s1}$. The adjustment
of the hardware system for filter-bank sampling is incremental with
no need to rebuild the whole system from scratch.

\bibliographystyle{IEEEtran} \bibliographystyle{IEEEtran} \bibliographystyle{IEEEtran}

\section*{Acknowledgement}

\addcontentsline{toc}{section}{Acknowledgment}

This work was supported by the NSF Center for Science of Information,
the Interconnect Focus Center of the Semiconductor Research Corporation,
and BSF Transformative Science Grant 2010505.\bibliographystyle{IEEEtran}
\bibliography{bibfileNonuniform}

\end{document}